\documentclass[prb,showpacs,twocolumn,floatfix]{revtex4}

\usepackage{amsmath,amsfonts}
\usepackage{graphicx}
\usepackage{color}
\usepackage{enumerate}
    
\newcommand{\e}{\mathrm{e}}
\newcommand{\up}{\uparrow}
\newcommand{\dw}{\downarrow}
\newcommand{\mean}[1]{\langle #1 \rangle}
\newcommand{\bx}{\mathbf{x}}
\newcommand{\by}{\mathbf{y}}
\newcommand{\bEta}{\boldsymbol{\eta}}
\newcommand{\bsigma}{\boldsymbol{\sigma}}

\date{\today}

\begin{document}

\title{Spin-selective Peierls transition in interacting one-dimensional conductors with spin-orbit interaction}

\author{Bernd Braunecker,$^1$ George I. Japaridze,$^{2,3}$ Jelena Klinovaja,$^{1}$ and Daniel Loss$^{1}$}
\affiliation{$^1$Department of Physics, University of Basel,
             Klingelbergstrasse 82, 4056 Basel, Switzerland}
\affiliation{$^2$Andronikashvili Institute of Physics, Tamarashvili 6, 0177 Tbilisi, Georgia} 
\affiliation{$^3$Ilia State University, Cholokashvili Ave. 3-5, 0162 Tbilisi, Georgia}

% ----------------------------------------------------------------------------

\begin{abstract}
Interacting one-dimensional conductors with Rashba spin-orbit coupling
are shown to exhibit a spin-selective Peierls-type transition into a mixed
spin-charge density wave state.
The transition leads to a gap for one-half of the conducting modes, which
is strongly enhanced by electron-electron interactions.
The other half of the modes remains in a strongly renormalized gapless state and 
conducts opposite spins in opposite directions, thus providing a perfect spin filter.
The transition is driven by magnetic field and by spin-orbit interactions.
As an example we show for semiconducting quantum wires and carbon nanotubes
that the gap induced
by weak magnetic fields or intrinsic spin-orbit interactions
can get renormalized by one order of magnitude up to 10-30 Kelvins.
\end{abstract}

% ----------------------------------------------------------------------------

\maketitle

% ----------------------------------------------------------------------------

\section{Introduction} 

Electron-electron (e-e) interactions play a
central role in determining the physical properties of
low-dimensional electron conductors. They lead to interesting
correlated-electron physics but also provide a handle
to design systems with specific properties that cannot be reached
with noninteracting particles. In this paper we focus on
one-dimensional (1D) electron conductors with strong spin-orbit
interaction (SOI).
Such systems have spin-split bands with generally a crossing of two bands 
with opposite spin projection. 
Through an external magnetic field or intrinsic SOI 
the degeneracy at the crossing point can be lifted
and a gap $\Delta$ opens.
We show in this paper that in this situation e-e interactions
play a crucial role, which has not been investigated so far.
The interactions lead to a substantial enhancement of the gap
and strongly modify the nature of the remaining conducting modes. 
Underlying the strong response to interactions is the instability 
of any 1D conductor to the formation of charge- and spin-density waves.
A spatially modulated potential can make this instability dominant
and drive the system into an ordered density wave phase, known as 
the electronic Peierls transition.
We show below that the renormalization of $\Delta$
can be identified with a Peierls-type transition depending on spin and 
chirality, affecting  only one-half of the conducting modes but both spin 
components. For short, we refer to it as \emph{spin-selective Peierls transition}.

\begin{figure}
    \includegraphics[width=\columnwidth]{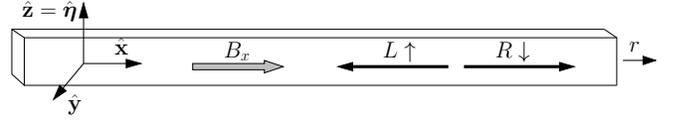}
    \caption{\label{fig:wire}Sketch of the quantum wire. The spin $z$ axis is
    chosen along the SOI axis $\bEta$.
	A magnetic field $B_x$ (if required) is applied perpendicularly to $\bEta$, for instance,
	along the spin $x$ axis.
	The position coordinate along the wire is denoted by $r$. 
	The labels $L \up$ and $R \dw$ indicate the spin-filtered conducting modes.
    }
\end{figure}

\begin{figure}
    \includegraphics[width=0.8\columnwidth]{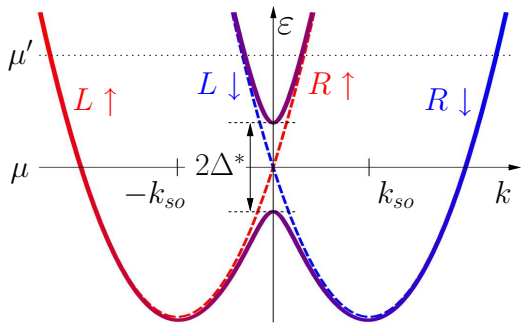}
    \caption{\label{fig:soi}
    Electron dispersion $\varepsilon_k$ for the 1D conductor.
    With the choice of spin axes as in Fig. \ref{fig:wire}, the $\sigma=\up, \dw$
    bands are shifted to the left and right by
    the wave vectors $\mp k_{so}$, respectively. Higher subbands are assumed to have no influence.
    Close to the chemical potential $\mu$ the modes are classified into left- ($L$)
    and right- ($R$) movers.
    The $L\dw$ and $R\up$ branches cross at $k=0$, and $\mu$ passes through 
	the crossing point by tuning the density to $k_F = k_{so}$.
 	A gap $2\Delta$ opens at $k=0$ through a magnetic field $B_x$
	or through intrinsic SOI, and is strongly enhanced by e-e interactions 
	to $2 \Delta^*$. The dashed lines correspond to $\Delta=0$.
	Through the application of a spiral magnetic field a gap can also open
	at any higher chemical potential $\mu'$.
    }
\end{figure}

A sketch of the conductor is shown in Fig. \ref{fig:wire}. 
The main effect of SOI is illustrated in Fig. \ref{fig:soi}
and consists in a spin-dependent shift of the electron dispersion 
by a momentum $\sigma k_{so}$, where $\sigma = \up,\dw = +,-$ is 
the spin polarization along an axis $\bEta$ determined by the SOI.
A uniform magnetic field perpendicular to $\bEta$ opens the gap $\Delta$
at the band crossing point at $k=0$. It was shown in previous work
\cite{streda:2003,pershin:2004,devillard:2005,zhang:2006,sanchez:2008,birkholz:2009}
that this leads to a reduced conductance and remaining spin-filtered 
conducting states.
This effect was very recently observed
in high-mobility GaAs/AlGaAs hole quantum wires.\cite{quay:2010}

We show here that e-e interactions substantially modify the physics 
within this gap. Indeed, tuning the chemical potential $\mu$ to the middle of $\Delta$ leads to 
the commensurability condition $k_F = k_{so}$ (with $k_F$ the Fermi momentum)
at which the e-e interactions have remarkable consequences.
They enhance $\Delta$ and for strong e-e interactions an enhancement by 
more than an order of magnitude is possible.
As an immediate consequence, the spin filter effect is stabilized,
removing the need of fine-tuning external parameters such as chemical
potential and magnetic field.
We show below that this renormalization can be mapped onto a Peierls-type
mechanism.
The interactions also modify the right- ($R$) and left- ($L$) moving modes
with momenta close to $\pm 2k_F$, which remain in a spin-filtered conducting
state, but form a strongly renormalized electron liquid.
Furthermore, we show that the same transition can be achieved without SOI through 
a spiral magnetic field as obtained, for instance, by placing 
nanomagnets near the conductor, or by the 
Overhauser field generated by nuclear spins through a self-ordering feedback 
mechanism due to their interaction with electrons.\cite{braunecker:2009a,braunecker:2009b}
For systems with SOI, a spiral magnetic field with wave number $2(k_F-k_{so})$
can also be used to obtain the renormalized gap at higher electron densities
$k_F > k_{so}$.
%

% ----------------------------------------------------------------------------

\section{Main physics}

To show the crucial importance of e-e interactions, let us consider a generic model
model for the 1D conductor, 
described by the Hamiltonian $H = H_{el} + H_{so} + H_\Delta$, where 
\begin{equation} \label{eq:H_el}
	H_{el} 
	= \sum_\sigma \int dr \
	\psi_\sigma^\dagger(r) \frac{(-i\hbar \partial_r)^2}{2m} \psi_\sigma(r)
	+ U,
\end{equation}
is the electron Hamiltonian,
\begin{equation} \label{eq:soi}
	H_{so} 
	= \sum_{\sigma\sigma'} \int dr \
	\psi^\dagger_{\sigma}(r)
	(\bEta \cdot \bsigma)_{\sigma \sigma'} (-i\hbar \partial_r) \psi_{\sigma'}(r),
\end{equation}
is the SOI, and 
\begin{equation} \label{eq:H_Delta}
	H_\Delta 
	= - \Delta \sum_{\sigma,\sigma'} \int dr \ 
	\psi_{\sigma}^\dagger(r) (\sigma_x)_{\sigma\sigma'} \psi_{\sigma'}(r),
\end{equation}
describes the gap.
In these Hamiltonians $\psi_\sigma(r)$ is the electron
operator for spin $\sigma$ at position $r$, $m$ is
the band mass, $U$ is a general e-e interaction
term, and $\bsigma = (\sigma_x,\sigma_y,\sigma_z)$ is the vector formed
by the Pauli matrices.
The vector $\bEta$ determines the amplitude and direction of the SOI
and depends on the material and confinement of the 1D system
(for instance, for Rashba SOI it is usually perpendicular to the 
wire axis). 
We choose the spin $z$ direction such 
that $\bEta \cdot \bsigma = \eta \sigma_z$. The gap $\Delta$ is typically
created by a uniform magnetic field $B_x$ applied perpendicularly to 
$\bEta$ along the spin $x$ direction but it can appear also through 
intrinsic SOI as in carbon nanotubes\cite{izumida:2009,jeong:2009} 
(see discussion in Sec. \ref{sec:experiments}). 
In the case of magnetic field,
$\Delta = B_x g\mu_B/2$, with $g$ the Land\'e $g$-factor and
$\mu_B$ the Bohr magneton. 
Further SOI and orbital coupling to the magnetic field
may be included without changing the described physics.

In the absence of e-e interactions ($U=0$) the electron dispersion can easily be
calculated and is shown in Fig. \ref{fig:soi} without (dashed lines)
and with (solid lines) the gap. 
The main effect of the SOI is a spin-dependent shift of
the single-particle dispersions by momentum $k_{so} = m \eta/\hbar$.
The spin-flip scattering of Eq. \eqref{eq:H_Delta} opens the gap $\Delta$ 
at $k=0$.

\begin{figure}
	\includegraphics[width=0.8\columnwidth]{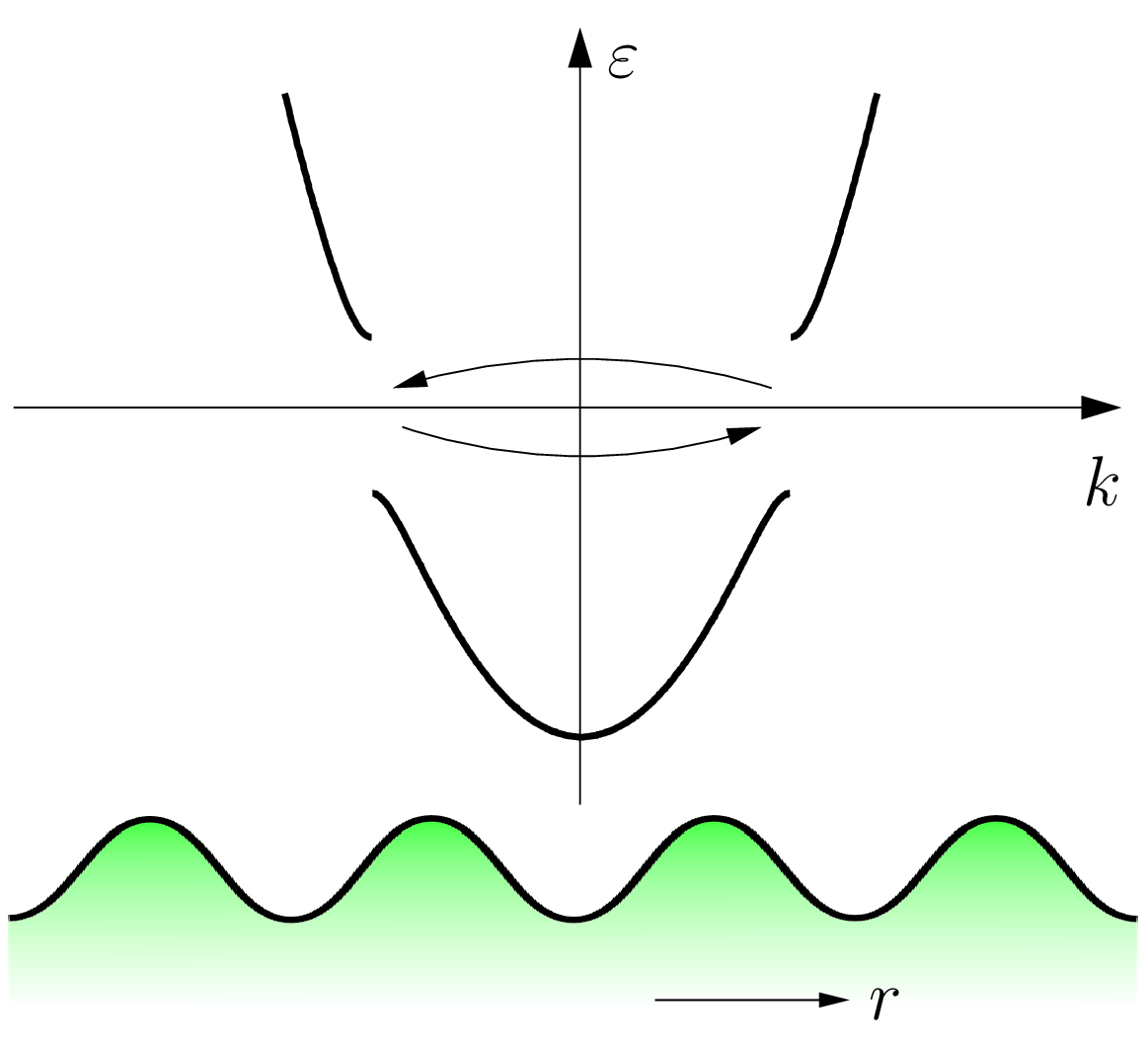}
	\caption{\label{fig:peierls}%
	Illustration of the electronic Peierls transition in a one-dimensional conductor.
	The scattering of electrons on an external periodic potential
	(represented in real space below the band structure) causes 
	hybridization between the states near $\pm k_F$ if the potential
	has the wave number $2k_F$. As a consequence a gap opens at $\pm k_F$
	and the system becomes insulating.
	}
\end{figure}

\begin{figure}
	\includegraphics[width=0.8\columnwidth]{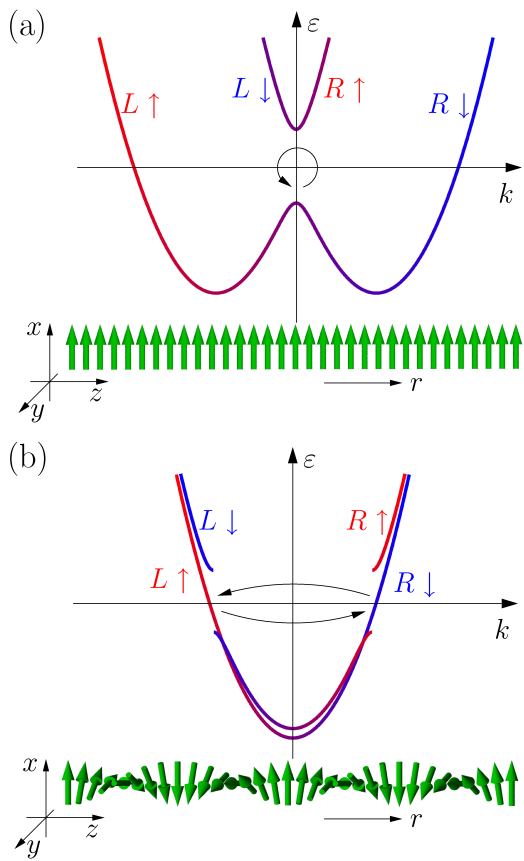}
	\caption{\label{fig:shift}%
	Illustration of the connection to the Peierls mechanism. 
	(a) Band structure of the SOI shifted bands as in Fig. \ref{fig:soi}.
	The arrows below the band structure represent the uniform magnetic
	field $B_x$ in real space. Spin-flip scattering on $B_x$
	(indicated by the circular arrow) hybridizes the modes at $k=0$ and
	opens the gap $\Delta \propto B_x$.
	(b) The same band structure in the gauge transformed basis
	$\psi_\sigma(r) \to \e^{i \sigma k_{so} r} \psi_\sigma(r)$. 
	The spin-split bands overlap, and the uniform magnetic field
	turns into a spiral field in the spin $x,y$ plane with wave number
	$2k_{so}$. The scattering on the magnetic field turns into a 
	spin-flip scattering on the periodic external potential with momentum
	transfer $\pm 2k_{so}$, as indicated by the arrows, and 
	leads to the hybridization between the modes $L\dw$ and $R\up$ 
	at $\pm k_{so}$.
	At $k_F = k_{so}$ this is equivalent to the Peierls 
	mechanism shown in Fig. \ref{fig:peierls},
	yet affects only one half of the electron modes. 
	Electron-electron interactions strongly renormalize $\Delta$
	and modify the properties of the remaining gapless states.
	For clarity of the presentation the spin axes are chosen here
	differently than in Fig. \ref{fig:wire}.
	}
\end{figure}

The opening of this gap is equivalent to a Peierls transition. 
The latter refers to the formation of an electron density wave 
in a 1D electron system in the presence of an external potential with 
a spatial periodicity commensurate with $2k_F$. 
Scattering on this potential with momentum transfer $\pm 2k_F$ 
leads to mixing between the modes with momenta close to $\pm k_F$
and turns the system into an insulator (see Fig. \ref{fig:peierls}).
An equivalent situation is obtained in the present system by considering
the spin-dependent gauge transformation
$\psi_\sigma(r) \to \e^{i \sigma k_{so} r} \psi_\sigma(r)$,
which eliminates the $\sigma k_{so}$ shifts such that
$H_0+H_{so} \to H_0$
(see Fig. \ref{fig:shift}).
The uniform $\Delta$, however, turns into the spiral field,
$\Delta(r) = \Delta [ \cos(2k_{so} r) \hat{\bx} - \sin(2k_{so} r) \hat{\by}]$,
fully equivalent to a spiral magnetic field in a system without SOI,
where $\hat{\bx},\hat{\by}$ are the unit vectors in the spin $x,y$ directions.
At commensurability $k_F = k_{so}$ this spiral plays the same role as the 
periodic potential for the regular Peierls instability.
Yet it is spin selective because the helicity breaks the symmetry between
the spin directions and so the gap 
$\Delta$ opens only between the modes $R \up$ and $L \dw$.

This identification with a Peierls-type mechanism tells us that e-e interactions 
substantially enhance $\Delta$.
We first visualize this with a simple mean-field picture,
equivalent to the Stoner argument, and give quantitative results afterward.
Let us assume that $\Delta$ is caused by a uniform field $B_x$. 
This field causes a paramagnetic partial spin polarization
$\mean{S_x}$ by the relation $\mean{S_x} = -\chi B_x$ with $\chi>0$
a susceptibility. We assume for illustration that the e-e interactions can be represented by a
simple interaction density of the type $U n_\sigma n_{\sigma'}$,
where $n_\sigma$ is the density of $\sigma$ electrons.
Important is the mean-field exchange coupling
$- 4 U \mean{S_x} S_x$ (using $\mean{S_{y,z}}=0$) with
$S_{x,y,z}$ the components of the electron spin density.
This exchange term acts as an effective magnetic field and leads to 
$\mean{S_x} = - \chi [B_x - 4 U \mean{S_x}] = - \chi B_x^*$
with the renormalized effective magnetic field $B_x^* = B_x / (1-4 \chi U)$
and so to the renormalized gap $\Delta^* = \Delta / (1-4 \chi U)$.

While this mean-field argument qualitatively shows that e-e
interactions can strongly enhance $\Delta$, it is quantitatively inaccurate for 1D systems
where the physics is dominated by fluctuations. To overcome this
limitation we use the Luttinger liquid (LL) theory.\cite{LL} 
Much attention was given recently to the connection between LL and SOI physics,
\cite{moroz:2000a,moroz:2000b,governale:2002,
iucci:2003,yu:2004,pereira:2005,li:2005,gritsev:2005,cheng:2007,sun:2007,gangadharaiah:2008,
schulz:2009,japaridze:2009,schulz:2010}
yet the addressed regimes are far from the regime
considered in this work, and the resulting physics is 
different. 
The LL theory is exact for
a linear electron dispersion, and band curvature can lead to
modified physics.\cite{LL_curv_1,LL_curv_2,LL_curv_3}
However, since the instability to 
density wave order is inherent in any 1D conductor, we expect that the 
Peierls-type renormalization persists for general situations, and we use the LL theory 
for quantitative predictions.

The physics of a LL in a spiral magnetic field 
was studied in detail in Refs. \onlinecite{braunecker:2009a} and \onlinecite{braunecker:2009b},
where it was shown that a Hamiltonian of the type of Eq. \eqref{eq:H_Delta}
can be written as
\begin{align} \label{eq:H_Delta_cos}
    H_\Delta
    = - \int dr &\frac{\Delta}{\pi a}
    \Bigl[
        \cos\bigl(\phi_{R\up}+\phi_{L\dw} + 2 (k_{so}-k_{F})r \bigr)
\nonumber\\
        &+
        \cos\bigl(\phi_{R\dw}+\phi_{L\up} + 2 (k_{so}+k_{F}) r\bigr)
    \Bigr],
\end{align}
where $a$ is the lattice constant
and $\phi_{\ell\sigma}$, for $\ell = L,R$, are boson fields such that 
$\partial_r \phi_{\ell\sigma}$ describe electron density fluctuations 
about the 4 Fermi points of the branches
labeled in Fig. \ref{fig:soi} (dashed lines).
In terms of the standard boson fields\cite{LL} for charge
($\phi_c, \theta_c$) and spin fluctuations ($\phi_s, \theta_s$)
we have
$\phi_{R\up}+\phi_{L\dw} = \sqrt{2}(\phi_c - \theta_s)$
and $\phi_{R\dw}+\phi_{L\up} = \sqrt{2}(\phi_c+\theta_s)$. 
The first cosine term in Eq. \eqref{eq:H_Delta_cos} describes the hybridization between the two
crossing branches. At $k_F = k_{so}$ it becomes
nonoscillating and relevant in the renormalization group (RG) sense. 
The second cosine, expressing scattering between the $R\dw$ and $L\up$ branches at $\pm 2k_F$,
then strongly oscillates (irrelevant for the RG) and can be neglected 
for systems of length $L \gg \pi/4k_F$.
The e-e interactions strongly enhance the amplitude of the relevant term,
and it was shown in Refs. \onlinecite{braunecker:2009a} and \onlinecite{braunecker:2009b} that 
the resulting renormalized gap is given by 
\begin{equation} \label{eq:Delta_Z*}
    \Delta^* = \Delta (\xi/a)^{1-\kappa},
\end{equation}
with $\kappa = (K_c+K_s^{-1})/2$ and $\xi$ the correlation length of the gapped modes.
Here $K_{c,s}$ are the LL parameters for charge and spin fluctuations, 
which fully encode the interaction $U$.\cite{LL}
For a noninteracting system $K_c=K_s=1$ and $\Delta^* = \Delta$. 
Repulsive e-e interactions
lead to $0 < K_c < 1$.  In the absence of spin SU(2) breaking
interactions other than Eq. \eqref{eq:soi} we have $K_s=1$.
Otherwise $K_s \neq 1$, and the interplay with the SOI can lead to
interesting spin-density-wave phases.\cite{gritsev:2005,sun:2007,gangadharaiah:2008}
At commensurability $k_{so}=k_F$, however, the nonoscillating term 
in Eq. \eqref{eq:H_Delta_cos} is much more relevant and we can neglect such
processes.
For an infinite system at zero temperature $\xi$ is given by\cite{braunecker:2009b}
$\xi=\xi_\infty = (\pi \hbar v_F/\Delta a)^{1/(2-\kappa)}$ 
with $v_F$ the slope of the dispersion at $k_F=k_{so}$.
In a realistic system $\xi$ is furthermore limited by
the system length $L$ and the thermal length $\lambda_T = \hbar v_F/k_B T$
(with $k_B$ the Boltzmann constant) such that,
with an order 1 uncertainty, $\xi = \min\{L, \lambda_T,\xi_\infty\}$.
For the experimentally available systems considered in Sec. \ref{sec:experiments}
we have $\xi = \xi_\infty$ except at very low fields below some millitesla.
The gap is stable upon a slight detuning $\Delta k = |k_F - k_{so}|$ 
from commensurability as long as $\xi \ll \pi / \Delta k$.\cite{LL}

% ----------------------------------------------------------------------------

\section{Consequences}

For temperatures $k_B T < \Delta^*$ the combination
$\phi_{R\up}+\phi_{L\dw}$ is pinned
to a minimum of the relevant cosine term in Eq. \eqref{eq:H_Delta_cos}.
This leads to a spin polarization in $x$ direction, which we estimate as\cite{braunecker:2009b}
$\mean{S_x} \equiv \mean{\psi_{L\dw}^\dagger \psi_{R\up}} = \mean{\psi_{R\up}^\dagger \psi_{L\dw}}
\sim \Delta^* /2E_F$ with $E_F = \hbar v_F k_F/2$. 
This is essentially the mean-field result
from above (with $\chi \sim 1/E_F$) but with the corrected enhancement factor
$(\xi/a)^{1-\kappa}$.
Since $\xi$ depends on $\Delta$ 
(except for very small $\Delta$ where $\xi \neq \xi_\infty$)
the response is generally nonlinear.
Let $M_x = (\mu_B g/2) \mean{S_x}$ be the magnetization of the system. 
For $\Delta \propto B_x$, we find with $\mean{S_x} \propto \Delta^* \propto g$ that 
$M_x \propto g^2 (\xi/a)^{1-\kappa}$. Hence we can interpret the enhancement also as a 
$B_x$-dependent $g$-factor renormalization, $g \to g (\xi/a)^{(1-\kappa)/2}$,
yet only for the gapped modes.

The ungapped modes $L\up$ and $R\dw$ remain in a LL state describing the
low-energy physics with strongly modified parameters.
This can be captured by writing the corresponding electron operators as
\cite{braunecker:2009b,LL} $\psi_{L\up} \propto \e^{-i(\phi + \theta)}$
and $\psi_{R\dw} \propto \e^{i (\phi-\theta)}$, where we have used
$\phi_c = \theta_s +$ const. The LL theory is then described by the 
Hamiltonian\cite{braunecker:2009b} 
$H = \hbar v  \int (dr/2\pi) \left[ K^{-1} (\partial_r \phi)^2 + K (\partial_r \theta)^2 \right]$,
with $K = 2K_c/(1+K_sK_c)$ and 
$v = v_F (K_c^{-1} + K_c)/(1+ K_c K_s)$.
The gapless modes preserve the $L\up$ and $R\dw$
character, and the conductor acts as a perfect spin filter for injected
electrons.
Indeed, local charge fluctuations $\rho_c$ and spin current $j_s$ are identical,
$\rho_c(r) = j_s(r) = -\partial_r \phi(r)/\pi$, showing that any charge
fluctuation transports spin through the system.
This also implies that backscattering without a spin flip is strongly
suppressed, and that the system is stable against nonmagnetic backscatterers
such as impurities or disorder.

Replacing the uniform magnetic field $B_x$ by the spiral magnetic field  
$B_{xy}(r) = B_{xy} [ \cos(4k_{so}r) \hat{\bx} + \sin(4k_{so}r) \hat{\by}]$
(if feasible) exchanges the relevance of the cosine
terms in Eq. \eqref{eq:H_Delta_cos} by switching
$(k_{so} + k_F) \leftrightarrow (k_{so}- k_F)$. Consequently a gap opens
for the modes at $k \approx \pm 2k_F$ while the modes at $k\approx 0$ remain unaffected. 
The wave packets for the $k\approx 0$ states extend over the 
whole system and we expect them to be only weakly affected by local scatterers
such as impurities. More important are e-e interactions such as $k=0$ umklapp scattering, 
which can lead to the opening of an additional gap, clearly distinguishable from $\Delta^*$ 
by its dependence on system properties. 

Different spiral magnetic fields allow to obtain the gap $\Delta^*$
already at higher electron densities (position of $\mu'$ in Fig. \ref{fig:soi}). 
If $k_F = k_{so} + \Delta k$, the application of 
$B_{xy}(r) = B_{xy} [ \cos(2 \Delta k r) \hat{\bx} - \sin(2 \Delta k r) \hat{\by}]$
eliminates the oscillation of the first cosine term in Eq. \eqref{eq:H_Delta_cos}
and opens the gap.
Since $k_{so} \neq 0$ the transition is obtained with a spiral field of longer
wavelength than in the absence of SOI.

% ----------------------------------------------------------------------------

\section{Experimental realizations}
\label{sec:experiments}

Candidates for this physics are 
InAs nanowires\cite{fasth:2007,nadj-perge:2010} and AlGaAs quantum wires,\cite{quay:2010}
where $\Delta$ is induced by $B_x$; or single-wall carbon nanotubes, where $\Delta$ 
appears intrinsically from SOI.\cite{izumida:2009,jeong:2009}
The latter are interesting candidates as they can be produced at high purity 
and have strong e-e interactions.\cite{cnt_1,cnt_2,cnt_3} Importantly, the nanotube curvature
induces intrinsic SOI,\cite{izumida:2009,jeong:2009} which leads to a gap at the band 
crossing of metallic nanotubes. However, due to the hollow nature of the nanotubes, 
the spin-dependent shift $k_{so}$ is absent. The latter can be induced by Rashba SOI 
with an electric field across the nanotube. 
Following Ref. \onlinecite{min:2006}, we estimate that a splitting
such that a band structure as in Fig. \ref{fig:soi} can be resolved requires an electric
field on the order of 0.1--1 V/nm.\cite{klinovaja}
The effect of a transverse electric field on the band structure was investigated 
with perturbation theory in Ref. \onlinecite{novikov:2002}.
At such strong fields, however, the perturbative treatment must be replaced
by a full incorporation into the Bloch theory. The final Dirac theory remains very 
similar and SOI has the same effect. In particular, there remains the intrinsic band 
gap on the order of $\Delta \sim 40$ $\mu$eV (tunable by tube radius).\cite{izumida:2009,jeong:2009}
Using\cite{braunecker:2009a,braunecker:2009b}
$a=2.46$ \AA, $K_c \approx 0.2$, and $K_s \approx 1$
(but SOI may further modify $K_s$, see also Ref. \onlinecite{schulz:2010}),
we obtain $\Delta^* \sim 1$ meV, enhanced by about a factor 30.
(Note that here in the same way as in Refs. \onlinecite{braunecker:2009a} and \onlinecite{braunecker:2009b} 
the spin-selective Peierls
transition leads to an effective decoupling of the two valleys at the $K$ and $K'$ points,
and the important LL parameter remains $K_c \approx 0.2$.)

\begin{figure}[h]
	\includegraphics[width=0.9\columnwidth]{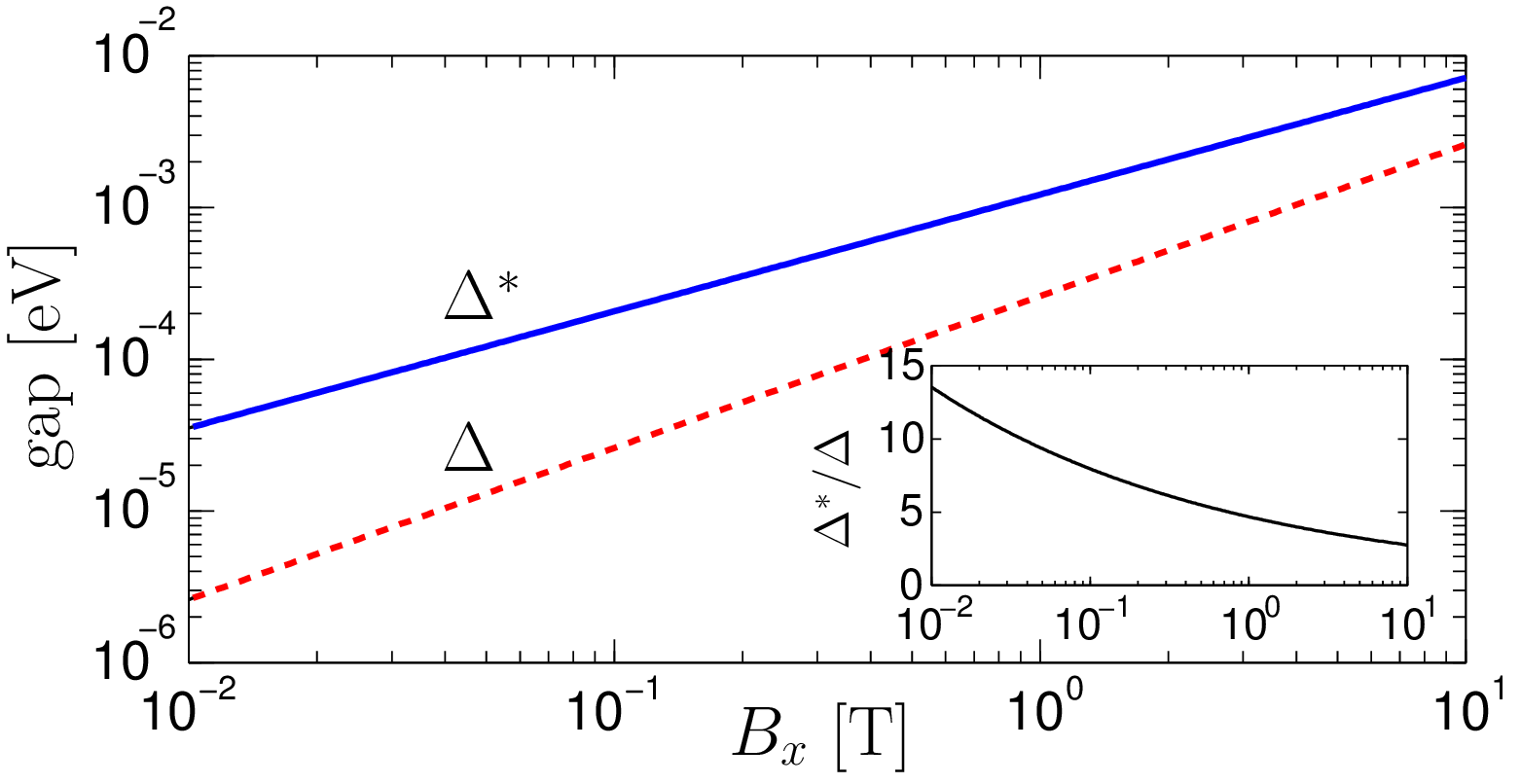}
	\caption{\label{fig:deltas}
	Gaps $\Delta$ and $\Delta^*$ as a function of magnetic field $B_x$ 
	for typical values of InAs quantum wires.
	The inset shows the enhancement $\Delta^*/\Delta$ for the same
	values of $B_x$.
	}
\end{figure}

In contrast, InAs nanowires have the spin-shift $k_{so}$ but require a magnetic field
to open the gap. 
For such wires it has been shown\cite{fasth:2007,nadj-perge:2010} that a SOI length 
of $\lambda_{so} \approx 130$ nm can be obtained, corresponding to 
$k_{so} =2\pi/\lambda_{so} = 5 \times 10^{7}$ m$^{-1}$.
Similar numbers are found in InAs quantum wells, which have
SOI strengths of\cite{zutic:2004,silsbee:2004,simmonds:2008b}
$\hbar\eta = (0.6-4) \times 10^{-11}$ eV m
corresponding to $k_{so} = 1.2 \times 10^{7}$ m$^{-1}$ 
(using\cite{simmonds:2008} $m=0.040 m_e$, with $m_e$ the electron mass, $g = -9$, 
and a maximal $\eta$).
Comparable values have also been reported for AlGaAs quantum wires, \cite{quay:2010} yet 
in the hole doped regime, which is not explicitly addressed with this theory.
For such low densities it is reasonable to assume $K_c \sim 0.4$
and $K_s = 1$. With $a = 6.06$~\AA\ we obtain the values shown in 
Fig. \ref{fig:deltas}. Notably an enhancement by more than a factor 10
is achieved at low fields.
%

% ----------------------------------------------------------------------------

\section{Conclusions}

We have shown that in 1D conductors with SOI the gap $\Delta$ opened by a magnetic field 
at the crossing point of the spin-split bands is substantially enhanced by e-e interactions.
In the local rest frame
of the electron spin $\Delta$ becomes a spiral field and assumes
the role of a spin-selective periodic potential that drives the system
through a Peierls-type transition. 
The interactions also renormalize the remaining gapless modes and strongly 
stabilize their spin-filter effect. 
Remarkably, it is possible to obtain the same effects even without a time-reversal
symmetry breaking magnetic field in carbon nanotubes, where a gap $\Delta$
exists intrinsically from SOI alone.

% ----------------------------------------------------------------------------

\begin{acknowledgments}
We thank M. J. Schmidt for helpful discussions.
B.B., J.K., and D.L. acknowledge the support by the Swiss NSF and NCCR Nanoscience (Basel),
and GIJ the support by the Georgian NSF under Grant No. ST/09-447. 
\end{acknowledgments}

% ----------------------------------------------------------------------------

% ----------------------------------------------------------------------------

\begin{thebibliography}{88}

\bibitem{streda:2003}
P. St\v{r}eda and P. \v{S}eba,
Phys. Rev. Lett. {\bf 90}, 256601 (2003).

\bibitem{pershin:2004}
Y. V. Pershin, J. A. Nesteroff, and V. Privman,
Phys. Rev. B {\bf 69}, 121306(R) (2004).

\bibitem{devillard:2005}
P. Devillard, A. Cr\'{e}pieux, K. I. Imura, and T. Martin,
Phys. Rev. B {\bf 72}, 041309(R) (2005).

\bibitem{zhang:2006}
S. Zhang, R. Liang, E. Zhang, L. Zhang, and Y. Liu,
Phys. Rev. B {\bf 73}, 155316 (2006).

\bibitem{sanchez:2008}
D. S\'{a}nchez, L. Serra, and M.-S. Choi,
Phys. Rev. B {\bf 77}, 035315 (2008).

\bibitem{birkholz:2009}
J. E. Birkholz and V. Meden,
Phys. Rev. B {\bf 79}, 085420 (2009).

\bibitem{quay:2010}
C. H. L. Quay, T. L. Hughes, J. A. Sulpizio, L. N. Pfeiffer, 
K. W. Baldwin, K. W. West, D. Goldhaber-Gordon, and R. de Picciotto,
Nat. Phys. {\bf 6}, 336 (2010).
 
\bibitem{braunecker:2009a}
B. Braunecker, P. Simon, and D. Loss,
Phys. Rev. Lett. {\bf 102}, 116403 (2009).

\bibitem{braunecker:2009b}
B. Braunecker, P. Simon, and D. Loss,
Phys. Rev. B {\bf 80}, 165119 (2009).

\bibitem{izumida:2009}
W. Izumida, K. Sato, and R. Saito,
J. Phys. Soc. Jpn. {\bf 78}, 074707 (2009).

\bibitem{jeong:2009}
J.-S. Jeong and H.-W. Lee, 
Phys. Rev. B {\bf 80}, 075409 (2009).

\bibitem{LL}
A. O. Gogolin, A. A. Nersesyan, and A. M. Tsvelik,
\emph{Bosonization and Strongly Correlated Systems},
(Cambridge University Press, Cambridge, 1998);
T. Giamarchi,
\emph{Quantum Physics in One Dimension},
(Oxford University Press, Oxford, 2004).

\bibitem{moroz:2000a} 
A. V. Moroz, K. V. Samokhin, and C. H. W. Barnes, 
Phys. Rev. Lett. {\bf 84}, 4164 (2000).

\bibitem{moroz:2000b} 
A. V. Moroz, K. V. Samokhin, and C. H. W. Barnes, 
Phys. Rev. B {\bf 62}, 16900 (2000).

\bibitem{governale:2002}
M. Governale and U. Z\"{u}licke,
Phys. Rev. B {\bf 66}, 073311 (2002).

\bibitem{iucci:2003}
A. Iucci,
Phys. Rev. B {\bf 68}, 075107 (2003).

\bibitem{yu:2004}
Y. Yu, Y. Wen, J. Li, Z. Su, and S. T. Chui,
Phys. Rev. B {\bf 69}, 153307 (2004).

\bibitem{pereira:2005}
R. G. Pereira and E. Miranda,
Phys. Rev. B {\bf 71}, 085318 (2005).

\bibitem{li:2005}
Y.-X. Li, Y. Guo, and B.-Z. Li,
Phys. Rev. B {\bf 72}, 075321 (2005).

\bibitem{gritsev:2005}
V. Gritsev, G. I. Japaridze, M. Pletyukhov, and D. Baeriswyl,
Phys. Rev. Lett. {\bf 94}, 137207 (2005).

\bibitem{cheng:2007}
F. Cheng and G. Zhou,
J. Phys.: Condens. Matter {\bf 19}, 136215 (2007).

\bibitem{sun:2007}
J. Sun, S. Gangadharaiah, and O. A. Starykh,
Phys. Rev. Lett. {\bf 98}, 126408 (2007).

\bibitem{gangadharaiah:2008}
S. Gangadharaiah, J. Sun, and O. A. Starykh,
Phys. Rev. B {\bf 78}, 054436 (2008).

\bibitem{schulz:2009}
A. Schulz, A. De Martino, P. Ingenhoven, and R. Egger,
Phys. Rev. B {\bf 79}, 205432 (2009).

\bibitem{japaridze:2009}
G. I. Japaridze, H. Johannesson, and A. Ferraz,
Phys. Rev. B {\bf 80}, 041308(R) (2009).

\bibitem{schulz:2010}
A. Schulz, A. De Martino, and R. Egger,
Phys. Rev. B {\bf 82}, 033407 (2010)

\bibitem{LL_curv_1}
M. Khodas, M. Pustilnik, A. Kamenev, and L. I. Glazman,
Phys. Rev. B {\bf 76}, 155402 (2007).

\bibitem{LL_curv_2}
A. Imambekov and L. I. Glazman,
Science {\bf 323}, 228 (2009).

\bibitem{LL_curv_3}
A. Imambekov and L. I. Glazman,
Phys. Rev. Lett. {\bf 102}, 126405 (2009).

\bibitem{fasth:2007}
C. Fasth, A. Fuhrer, L. Samuelson, V. N. Golovach, and D. Loss,
Phys. Rev. Lett. {\bf 98}, 266801 (2007).

\bibitem{nadj-perge:2010}
S. Nadj-Perge, S. M. Frolov, J. W. W. van Tilburg, J. Danon, 
Y. V. Nazarov, R. Algra, E. P. A. M. Bakkers, and L. P. Kouwenhoven
Phys. Rev. B {\bf 81} 201305(R) (2010).

\bibitem{cnt_1}
C. Sch\"{o}nenberger,
Semicond. Sci. Technol. {\bf 21}, S1 (2006).

\bibitem{cnt_2}
A. K. H\"{u}ttel, G. A. Steele, B. Witkamp, M. Poot, L. P. Kouwenhoven, and H. S. J. van der Zant, 
Nano Lett. {\bf 9}, 2547 (2009).

\bibitem{cnt_3}
H. O. H. Churchill, A. J. Bestwick, J. W. Harlow, F. Kuemmeth, D. Marcos, 
C. H. Stwertka, S. K. Watson, and C. M. Marcus,
Nat. Phys. {\bf 5}, 321 (2009).

\bibitem{min:2006}
H. Min, J. E. Hill, N. A. Sinitsyn, B. R. Sahu, L. Kleinman, and A. H. MacDonald,
Phys. Rev. B {\bf 74}, 165310 (2006).

\bibitem{klinovaja}
J. Klinovaja, M. J. Schmidt, B. Braunecker, and D. Loss, 
in preparation.

\bibitem{novikov:2002}
D. S. Novikov and L. S. Levitov, 
Phys. Rev. Lett. {\bf 96}, 036402 (2006);
arXiv:cond-mat/0204499.

\bibitem{zutic:2004}
I. \v{Z}uti\'{c}, J. Fabian, S. Das Sarma,
Rev. Mod. Phys. {\bf 76}, 323 (2004).

\bibitem{silsbee:2004}
R. H. Silsbee,
J. Phys.: Condens. Matter {\bf 16}, R179 (2004).

\bibitem{simmonds:2008b}
P. J. Simmonds, S. N. Holmes, H. E. Beere, and D. A. Ritchie
J. Appl. Phys. {\bf 103}, 124506 (2008).

\bibitem{simmonds:2008}
P. J. Simmonds, F. Sfigakis, H. E. Beere, D. A. Ritchie, M. Pepper, D. Anderson, and G. A. C. Jones,
Appl. Phys. Lett. {\bf 92}, 152108 (2008).

 
\end{thebibliography}
\end{document}